\documentclass[aps,pra,twocolumn,groupedaddress,showkeys]{revtex4}
\usepackage{graphicx}% Include figure files
\usepackage{dcolumn}% Align table columns on decimal point
\usepackage{bm}% bold math
\usepackage{amsmath}
\usepackage{amssymb}
\usepackage{color} % \textcolor[rgb]{0,0,1}{text}
\usepackage{slashed}

\begin{document}

\title{Passive quantum measurement: \\
Arrival time, quantum Zeno effect and gambler's fallacy}

\author{Tajron Juri\'c}\email{tjuric@irb.hr}
\author{Hrvoje Nikoli\'c}\email{hnikolic@irb.hr}
\affiliation{Theoretical Physics Division, Rudjer Bo\v{s}kovi\'{c} Institute, P.O.B. 180, HR-10002 Zagreb, Croatia.}

\begin{abstract}
Classical measurements are passive, in the sense that they do not affect the physical properties 
of the measured system. Normally, quantum measurements are not passive in that sense.
In the infinite dimensional Hilbert space, however, we find 
that quantum projective measurement can be passive
in a way which is impossible in finite dimensional Hilbert spaces.
Specifically, we find that expectation value of a hermitian Hamiltonian can have an imaginary part
in the infinite dimensional Hilbert space and that such an imaginary part implies a possibility 
to avoid quantum Zeno effect,
which can physically be realized in quantum arrival experiments. The avoidance of quantum Zeno effect can also
be understood as avoidance of a quantum version of gambler's fallacy, leading to the notion of passive quantum  
measurement that updates information about the physical system without affecting its physical properties.  
The arrival time probability distribution of a particle is found to be given by the flux of the probability current.
Possible negative fluxes correspond to regimes at which there is no arrival at all,  
physically understood as regimes at which the particle departs rather than arrives. 
\end{abstract}


\maketitle

\section{Introduction} 
 
The Hilbert space in quantum mechanics is, strictly speaking, 
an {\em infinite} dimensional vector space,
because it deals with square-integrable wave functions. 
For many purposes, however, one can approximate it with a finite dimensional Hilbert space,
for example when one computes wave functions numerically on a computer, or when one 
restricts attention to intrinsically finite dimensional concepts such as spins and qubits.

The Dirac bra-ket formalism \cite{dirac} allows to treat infinite dimensional Hilbert spaces 
by heuristic rules that don't differ much from the matrix formalism for finite dimensional Hilbert spaces.
For infinite dimensional Hilbert spaces such a formalism is not mathematically rigorous, 
but in practice it works fine -- except when it doesn't!
In some cases the na\"{i}ve Dirac formalism leads to various inconsistencies, apparent paradoxes
and mathematical surprises \cite{gieres,gitman,juric} which require a more careful mathematical treatment.
In this paper we explore one such mathematical surprise and find that it has fundamental physical implications.
Under certain conditions, it turns out that
quantum measurement does not affect the measured system, thus behaving very much like 
a classical measurement. 

Measurement in classical physics is usually a passive measurement; it merely reveals a property of the measured system that 
preexisted before the measurement, without changing the measured system.
Typical quantum measurement, on the other hand, is very different because it changes the measured system.
There are various contextuality theorems \cite{KS,bell,peres,laloe} showing that quantum measurements, in general, 
do not merely reveal preexisting properties. Thus quantum measurements are generally viewed as physical processes
that play an active role in the process of measurement. Nevertheless we find that, in the infinite 
dimensional Hilbert space, quantum measurement can be passive in a way which is impossible in finite dimensional Hilbert spaces.   

At this point, some conceptual remarks are needed to avoid misunderstandings. 
Of course, both classical and quantum measurements involve an interaction 
with a measuring apparatus, and such an interaction always has an influence 
on the measured system. However, it is essential to keep in mind that 
this is {\em not} what we mean by measurement in this paper. 
Indeed, in quantum physics such an interaction corresponds to a unitary 
process and is often called {\em premeasurement} in the literature.
In principle, the effect of such an interaction can be arbitrarily small
but is never strictly zero, in both classical and quantum physics.
By measurement, however, in this paper we mean the {\em update of information} about the
measured system. In quantum physics, the update of information is described 
by a {\em non-unitary projection} of the wave function, and it is this projection
that makes quantum physics fundamentally different from classical physics.
The update of information has no influence at all on the measured system in classical 
physics, while in quantum physics it often has dramatic consequences.
It is the non-unitary projection, not the unitary interaction, 
that is responsible for the quantum uncertainty relations
(there is no single projector that projects to both a position and a momentum eigenstate),
and for the various contextuality theorems mentioned above.     
Finally, we stress that {\em explanation} of the projection 
is controversial and depends a lot 
on interpretation of quantum mechanics (such as Copenhagen, many-world, or Bohmian 
interpretation, to name a few), but to avoid controversy as much as possible,
in this paper we shall not try to {\em explain}
the projection; we shall only use it as a well-established rule that works in 
practice \cite{nielsen-chuang}.

Physically, the idea of passive quantum measurement is easy to understand. The simplest example
that discerns a typical active quantum measurement from a passive quantum measurement is a particle detector. 
When the detector clicks, we learn a new peace of information telling us that the particle is there, 
so we update the wave function by
``collapsing'' it with a projector $\pi$
\footnote{If the measurement is a non-projective POVM measurement \cite{nielsen-chuang}, 
e.g. because the 
detected particle is destroyed, it can still be described in terms of a projector in a bigger Hilbert space
that includes some ancilla degrees of freedom of the measuring apparatus \cite{peres,schumacher}.}. 
But when the detector does not click, we also learn a new peace of information; we learn that the particle is {\em not} there,
so we update the wave function again, this time with the projector $\bar{\pi}=1-\pi$. Intuitively, however, 
one expects that non-detection of the particle should not affect its physical properties. Non-detection
is expected to behave like a classical measurement, that can be interpreted as a passive update of information
without any active influence on the particle. The passive quantum measurement is a quantum measurement with such a 
classical behavior.   

Is passive quantum measurement consistent with principles of quantum mechanics?
It will turn out that its consistency is very nontrivial mathematically.
As we shall see, passive quantum measurement can be understood as a projective measurement that involves
a projector $\bar{\pi}$ with two seemingly contradictory properties
\begin{equation}\label{magic_intro}
 [H,\bar{\pi}] \neq 0, \;\;\;  \bar{\pi}[H,\bar{\pi}]=0 ,
\end{equation}
where $H$ is a hermitian Hamiltonian operator. It is not difficult to see that it is impossible
to simultaneously satisfy both properties in (\ref{magic_intro}) in any finite dimensional Hilbert space,
which will be shown in this paper explicitly. 
Surprisingly, however, recently it was found
in \cite{nik_tajron} 
%(in the context of quantum arrival time problem based on the approach initiated in \cite{jur-nik}) 
that (\ref{magic_intro}) can be satisfied in the infinite dimensional Hilbert space.  
In this paper we explore the mathematical origin and physical meaning of (\ref{magic_intro})
in more basic terms, leading us not only to a deeper understanding of the results in \cite{nik_tajron},
but also to a more general fundamental insight about quantum measurements.  

We note that the notion of passive quantum measurement  
is different from other known kinds of quantum measurements that avoid changing the measured system. 
Quantum non-demolition measurement \cite{braginsky,caves,shimizu} requires that the measured observable 
commutes with the Hamiltonian, which is not the case for passive quantum measurement.
Weak quantum measurement \cite{spin100,kofman,svensson,dressel} requires averaging over many repeated 
measurements to get a result, while passive quantum measurement reveals a result with a single measurement. 
The measurement proposed in \cite{nik-atelj} involves a macroscopic quantum system, such as a superconducting current,
while passive quantum measurement works for microscopic quantum systems.

The paper is organized as follows. In Sec.~\ref{SECexp}
we start from an elementary, yet surprising and unexpected, demonstration 
that, in the infinite dimensional Hilbert space, 
the expectation value of a hermitian Hamiltonian can have an imaginary part.
In Sec.~\ref{SECzeno} we explain how this imaginary part implies a possibility to avoid the 
quantum Zeno effect and how this possibility can actually be realized in quantum arrival experiments.
In Sec.~\ref{SECgambler} we rephrase the quantum Zeno effect as a quantum version of gambler's fallacy,
showing that avoiding quantum Zeno effect can be understood as a new kind of quantum measurement, called 
passive quantum measurement, which, similarly to classical measurements, updates information about the physical system 
without affecting its physical properties. 
In Sec.~\ref{SECarrival} we study the arrival time distribution in detail and find that the 
arrival probability density ${\cal P}_{\rm arr}(t)$ is equal to the flux of probability current through the outer boundary 
of the detector. Possible negative fluxes 
are interpreted physically as regimes in which  ${\cal P}_{\rm arr}(t)=0$ because the particle departs,
rather than arrives.     
A conclusion with prospects for further investigations is presented in Sec.~\ref{SECdisc}.

\section{Basic apparent paradox: Expectation value of a hermitian Hamiltonian can have a non-zero imaginary part}
\label{SECexp}

\subsection{The basic idea in physicist's language}

Consider the Hamiltonian of a free particle moving in one dimension
\begin{equation}\label{H}
 H=\frac{p^2}{2m}=\frac{-\hbar^2}{2m} \frac{d^2}{dx^2} .
\end{equation}
This Hamiltonian is hermitian so one expects that its expectation value should always be real 
\cite{schiff,sakurai,ballentine}. Nevertheless,
we shall demonstrate that its expectation value can have a non-zero imaginary part, 
provided that the wave function is slightly pathological. The precise meaning of ``slightly pathological'' will 
be explained later, but for now it suffices to say that such a wave function is 
somewhat untypical, but {\em not} unphysical.

All we require of our wave function $\psi(x)$ is that it has a support on a {\em finite open} interval 
$\bar{D}=(a,b)$ (the notation $\bar{D}$ anticipates that later it will 
be interpreted as the complement of the detector region $D$),
so that $\psi(x)$ in non-zero for $x\in \bar{D}$, but zero for $x\notin \bar{D}$. 
Hence the wave function satisfies
\begin{equation}\label{psi_chi}
 \psi(x)=\chi_{\bar{D}}(x)\psi(x) ,
\end{equation}
where  
\begin{equation}\label{pc2}
\chi_{\bar{D}}(x)\equiv\left\{ 
\begin{array}{ll}
 \displaystyle 1 & \;\; {\rm for} \;\; x\in \bar{D}  \\
 0 & \;\; {\rm for} \;\; x\notin \bar{D} 
\end{array}
\right.
\end{equation}
is the characteristic function associated with $\bar{D}$.  
We consider a wave function of this form at a {\em single} time,
typically at a time immediately after the measurement, thus serving as an ``initial'' condition for subsequent time 
evolution after the measurement. The wave function at this single time is non-vanishing only in $\bar{D}$ because we assume 
that the wave function collapsed to $\bar{D}$ due to measurement.
Thus the expectation value of $H$ 
at this single time
is given by
\begin{eqnarray}\label{expect}
\frac{2m}{\hbar^2}\langle\psi|H|\psi\rangle &=& -\int_{-\infty}^{\infty}dx\, \psi^*(x)\psi''(x)
%\nonumber \\
% &=& \textcolor[rgb]{1,0,0}{- \lim_{\epsilon\to 0^+}  \int_{a+\epsilon}^{b-\epsilon}dx\, \psi^*(x)\psi''(x)} ,
\nonumber \\
 &=& -  \int_{\bar{D}} dx\, \psi^*(x)\psi''(x) ,
\end{eqnarray}
where the primes denote derivatives with respect to $x$.
Notice that since $\bar{D}=(a,b)$ is {\em open}, the limits of the integration are  $a+\epsilon$ and 
$b-\epsilon$, so that the boundary points $a$ and $b$ are not included. 
More precisely, in general we have
\begin{equation}
\int_{\bar{D}} dx\, f(x)=\lim_{\epsilon\to 0^+}  \int_{a+\epsilon}^{b-\epsilon}dx\, f(x)
\end{equation}
for any measurable (generalized) function $f$.

After a partial integration, (\ref{expect}) can be written as
\begin{equation}\label{expect2}
 \frac{2m}{\hbar^2}\langle\psi|H|\psi\rangle = A+B,
\end{equation}
where
\begin{equation}\label{E}
 A= \int_{\bar{D}}  dx\, \psi'^*(x)\psi'(x) 
\end{equation}
is the usual ``bulk'' term, which is real and positive, while
\begin{eqnarray}\label{B}
 B &=& -\int_{\bar{D}} dx\, [\psi^*(x)\psi'(x)]'
\\
&=& \lim_{\epsilon\to 0^+}
[\psi^*(a+\epsilon)\psi'(a+\epsilon)-\psi^*(b-\epsilon)\psi'(b-\epsilon)]
\nonumber 
\end{eqnarray}
is the boundary term. 
The whole point is that {\em the boundary term $B$ does not necessarily need to be zero}.
In general it can be a complex number, implying that the expectation value of $H$ can have an imaginary part.
However, as we shall see in Sec.~\ref{SECproj},
the peculiar behaviour of $\psi(x)$ at the boundary implies that 
such a $\psi(x)$ is not in the domain of the Hamiltonian $H$, so the boundary effect 
cannot be seen by formally expanding $\psi(x)$ into the eigenfunctions of $H$. Physically, this means 
that such an imaginary value cannot appear as a result of measuring $H$.
As we shall see, the imaginary part of the expectation value of $H$ plays a role in a computation of time evolution 
generated by $H$, after a measurement of {\em another} observable that does not commute with $H$, 
and not in any measurement of $H$ itself, so the imaginary values are not numbers seen in experiments. 

To see how the boundary term can have a non-zero imaginary part, let us write the wave function in the polar form
$\psi(x)=R(x)e^{iS(x)}$, where $S(x)$ is real, while $R(x)$ is real and non-negative. Thus we can write
\begin{equation}\label{polar'}
 \psi^*(x)\psi'(x)=R(x)R'(x)+i|\psi(x)|^2 S'(x) ,
\end{equation}
where the first term is real but not necessarily positive, while the second term is imaginary. 
Hence (\ref{expect2}) can be written in the final form
\begin{equation}
 \frac{2m}{\hbar^2}\langle\psi|H|\psi\rangle = A + B_1 + i B_2 ,
\end{equation}
where $A$, $B_1$ and $B_2$ are real.
The term $B_1$ originates from the first term in (\ref{polar'}), which implies that the real part $A+B_1$
of the expectation value
of the Hamiltonian can be negative, despite the fact that (\ref{H}) is formally positive. It may have interesting 
physical applications, but in this paper we shall not further explore them. In this paper we shall 
only study the consequences of the imaginary term $iB_2$, which originates from the second term in (\ref{polar'}),
leading to
\begin{equation}\label{imdio}
 B_2= \lim_{\epsilon\to 0^+} [|\psi(a+\epsilon)|^2 S'(a+\epsilon)-|\psi(b-\epsilon)|^2 S'(b-\epsilon)].
\end{equation}
Therefore, whenever $\psi$ satisfies \eqref{psi_chi} and $S'\neq 0$, in general we can expect 
a non-vanishing imaginary part $B_2$ of the expectation value of the hermitian Hamiltonian \eqref{H}.
This is complementary to the work \cite{tumulka1,tumulka2}
which imposes the ``absorbing boundary condition'' that yields an a priori non-hermitian Hamiltonian,
while our approach allows 
non-vanishing imaginary part by leaving the wave function on the boundary completely arbitrary.

\subsection{The mathematical resolution of the apparent paradox}
\label{SECproj}

At first, the results above may look like a paradox. How can a positive hermitian operator have expectation value
with a negative real part and non-zero imaginary part? 

The resolution of the paradox can be understood in simple terms as follows.
Since $H$ is essentially a second-derivative operator on $\mathbb{R}$, 
an elementary proof that $H$ is hermitian involves 
a partial integration over whole real line $\mathbb{R}$ (not only over $\bar{D}$), 
which assumes that first and second derivatives of wave functions are well defined everywhere on $\mathbb{R}$.
Hence such a proof does not work for $\psi(x)$ obeying (\ref{psi_chi}), because such a wave function does not have 
well defined derivatives on the boundary of $\bar{D}$ (nevertheless, the limit in (\ref{B}) is well defined).  
Such a wave function does not belong to the class 
of wave functions for which $H\psi(x)$ is well defined everywhere,
i.e. such a $\psi(x)$ is not an element of the domain ${\cal D}(H)$ of the operator $H$. 
In the rest of this subsection we put this argument into a mathematically more sophisticated form.

Let us first review why a hermitian operator should have real expectation values. First of all, 
let us recall that a Hilbert space $\mathcal{H}$ is a vector space equipped with a scalar product with a property 
\begin{equation}\label{scal}
\langle\varphi|\psi\rangle=\langle\psi|\varphi\rangle^* \quad \forall \psi, \varphi\in \mathcal{H}.
\end{equation}
On the other hand, we say that an operator $H$ in Hilbert space is hermitian if 
\begin{equation}\label{her}
\langle\psi|H\varphi\rangle=\langle H\psi|\varphi\rangle .
\end{equation}
Now let us look at the expectation value of a hermitian operator, i.e. $\langle\psi|H|\psi\rangle$. 
Using the abbreviation $|H\psi\rangle\equiv |\xi\rangle$ we have
\begin{equation}\label{dokaz}
\langle\psi|H|\psi\rangle = \langle\psi|\xi\rangle
=\langle\xi|\psi\rangle
=\langle\psi|\xi\rangle^* ,
\end{equation}
where in the first equality we used the definition of the expectation value, 
in the second equality the hermiticity condition \eqref{her}, and in the last equality the property of the scalar product 
\eqref{scal}. Since the expectation value is a $c$-number, we conclude that the expectation value of a hermitian operator is real
\begin{equation}
\langle\psi|H|\psi\rangle\in\mathbb{R}
\end{equation}
and formally write $H=H^\dagger$. Even though this derivation is perfectly valid in a finite dimensional 
Hilbert space \footnote{Here the operator in question would be represented by a hermitian matrix. 
The derivation above may also be valid in infinite dimensional Hilbert space if  the operator in question is bounded, 
meaning that its domain is the whole Hilbert space.}, in the infinite dimensional Hilbert space it has a subtle flaw. 
Namely, there is no operator without its domain \cite{gieres,gitman,juric}. Operator is a rule of acting together with its domain. 
Changing the domain while keeping the same rule of acting completely changes the properties of an operator 
(like the spectrum, hermiticity, self-adjointness etc.) 
\footnote{For example, think about the momentum operator
given by the rule of acting via $p=-i\frac{d}{dx}$ and its corresponding eigenvalue problem on different domains. 
Different domains are specified by different boundary conditions such as Dirichlet, Neumann, Robin, periodic, etc.}. 
Most of the operators we encounter in quantum mechanics (like Hamiltonian, momentum, position, etc.) 
are unbounded operators and therefore cannot act on the whole Hilbert space, but rather on some subspace of 
it \footnote{Recall that most of these operators are represented by some differential operators, so for their domains 
we have to use functions which are not only square-integrable (that is, elements of $L^2(\mathbb{R})$), 
but also possess some additional properties such as 
differentiability, vanishing at infinity, obeying some boundary condition, etc.}. 
The domain of an unbounded operator is, at best, a dense subspace of the whole Hilbert space. 
Therefore, the hermiticity condition \eqref{her} should be reformulated as: 
{\em An operator $H: \mathcal{D}(H)\longrightarrow\mathcal{H}$ is hermitian if and only if}
\begin{equation}\label{HER}
\langle\psi|H\varphi\rangle=\langle H\psi|\varphi\rangle \quad \forall\  \psi, \varphi\in \mathcal{D}(H).
\end{equation}

Now we are ready to pinpoint the paradox, that is, understand how the imaginary part of the 
expectation value \eqref{imdio} is possible. Namely, the $\psi$'s in \eqref{imdio} are not in the domain of the hermitian 
Hamiltonian $H$. The operator $H$ is essentially a second-derivative operator, and for it to be well defined and hermitian 
(that is, satisfy \eqref{HER}), its domain $\mathcal{D}(H)$ has to be a dense subspace 
of the Hilbert space $L^2(\mathbb{R})$ 
that consists of certain ``well behaving'' functions. Namely, the maximal domain is given by 
\begin{equation}
\mathcal{D}_{\rm max}(H)=\left\{\psi\in L^2(\mathbb{R}) | H\psi\in L^2(\mathbb{R}) \right\}
\end{equation}
and on it the Hamiltonian \eqref{H} is a self-adjoint 
operator (that is, hermitian and $\mathcal{D}_{\rm max}(H)=\mathcal{D}_{\rm max}(H^\dagger)$). 
Of course, if we are not worried about self-adjointness, we could take for the domain the space of Schwartz functions 
(rapidly decreasing functions) $\mathcal{S}(\mathbb{R})$, or the space of smooth functions that vanish at infinity 
$\mathcal{C}_0^{\infty}(\mathbb{R})$ on which the Hamiltonian \eqref{H} is hermitian \footnote{Even more, 
it is essentially self-adjoint. For more details see \cite{reed, moretti}.}. 
The functions satisfying \eqref{psi_chi} are {\em not} elements of any of the three aforementioned  domains. 
The reason for this is simply because the characteristic function \eqref{pc2} is not weakly differentiable \cite{knjiga} 
and therefore 
any element $\psi$ of the Hilbert 
space that satisfies \eqref{psi_chi} can be written as $\psi=\chi_{\bar{D}}\varphi$,
which is clearly not an element of any $\mathcal{D}(H)$ that yields $H$ hermitian,
even when $\varphi$ is smooth.

However, this does not mean that such a wave function does not have any mathematical or physical sense. 
On the contrary, in this paper
we will see that such wave functions can have a useful and meaningful 
physical interpretation. 
But before going to the physical interpretation of such wave functions, let us look more closely at 
the mathematical meaning of the expectation value that leads to \eqref{imdio}. To say the least, 
the derivation of \eqref{imdio} is ill-defined because we are ``forcing'' a calculation where it is not applicable. 
We are basically feeding the Hamiltonian $H$ with an
input that is out of its domain. Can we mathematically 
justify this in a more rigorous way? In doing so, let us first rephrase the condition \eqref{psi_chi} 
in the Dirac notation 
\footnote{Dirac notation can be made rigorous 
by using rigged Hilbert space \cite{roberts,antoine,madrid}.} 
using the projector 
\begin{equation}\label{barpi}
 \bar{\pi}=\int_{\bar{D}} dx \,|x\rangle\langle x| , 
\end{equation}
so that the condition \eqref{psi_chi} boils down to 
\begin{equation}\label{psi_pi}
 |\psi\rangle = \bar{\pi}|\psi\rangle , 
\end{equation}
where we have used
\begin{eqnarray} 
& I=\displaystyle\int_{\mathbb{R}} dx \,|x\rangle\langle x| , &
\nonumber \\  
& \langle x|\bar{\pi}\psi\rangle=\chi_{\bar{D}}(x)\psi(x) , &
\nonumber \\
& \psi(x)\equiv\langle x|\psi\rangle=\langle x|\bar{\pi}\psi\rangle . &
\end{eqnarray}
Using \eqref{psi_pi}, the expectation value \eqref{expect} can be rewritten as
\begin{equation}
 \langle\psi|H|\psi\rangle = \langle\psi|\bar{\pi}H\bar{\pi}|\psi\rangle = \langle\psi|\overline{H}|\psi\rangle ,
\end{equation}
where
\begin{equation}\label{projected}
 \overline{H}\equiv \bar{\pi}H\bar{\pi} . 
\end{equation}
The operator $\overline{H}$ is the so called {\em projected} Hamiltonian and it was shown in \cite{nik_tajron} 
that it is {\em not hermitian},
which is another way to understand why the expectation value $\langle\psi|H|\psi\rangle$ can have an imaginary part. 
Furthermore, the projected Hamiltonian $\overline{H}$ has a unique property of generating the operator $V(t)$ 
defined as
\begin{equation}
V(t)=e^{-i\overline{H}t}\bar{\pi} .
\end{equation}
(Here, as in the rest of the paper, we work in units $\hbar=1$.) 
This operator governs 
the conditional time evolution induced by repeated projective measurements with outcomes $\bar{\pi}$ 
(for more details see Sec.~\ref{SECmore} or \cite{nik_tajron,jur-nik}). One can see that $V(t)$ satisfies
the following properties:
\begin{enumerate}
\item $V(t)$ is a bounded operator.
\item Contraction:   $\left\|V(t)\psi\right\|\leq\left\|\psi\right\|$.  
\item Semigroup:   $V(t)V(t^\prime)=V(t+t^\prime)$.  
\item Normalization:    $V(0)=\bar{\pi}=I_{\bar{\pi}\mathcal{H}}$.  
\item Strong continuity:   $\lim_{t\rightarrow 0}\left\|V(t)\psi-\psi\right\|=0_{\bar{\pi}\mathcal{H}}$. 
\item Differentiability:  $dV(t)/dt= -i\overline{H}V(t)$.
%$\frac{d V(t)}{dt}=-i\overline{H}V(t)$   
\end{enumerate}
Those properties mean
that $V(t)$ is a contraction operator that forms a strongly continuous semigroup when restricted to the 
projected Hilbert space $\bar{\pi}\mathcal{H}$. This is a consequence of the so called Hille--Yosida theorem 
\cite{Yosida,hille,engel}, 
which is a generalization of the famous Stone theorem for one parameter unitary groups \cite{stone,vonN}. 
Having this in mind, the expectation value $\langle\psi|H|\psi\rangle  = \langle\psi|\overline{H}|\psi\rangle$ 
can be rigorously defined as 
\begin{equation}
\langle\psi|\overline{H}|\psi\rangle\equiv i\left.\frac{d\langle\psi|V(t)|\psi\rangle}{dt}\right|_{t=0} .
\end{equation}

Note that our projected Hamiltonian $\overline{H}$ and the operator $V(t)$ can be regarded as extensions of the 
non-hermitian Hamiltonian and the contraction operator $W_t$ used in \cite{tumulka2}, where the 
absorbing boundary condition is used \cite{tumulka1}. However, due to the nature of the absorbing boundary condition 
(which is a Robin-type boundary condition), the approach in \cite{tumulka1,tumulka2} depends on one additional free parameter 
$\kappa$, the physical origin of which is not entirely clear, while in our approach there is no such free parameter.

% Similarly to other known paradoxes of that type \cite{gieres,gitman,juric}, 
% the resolution of the paradox consists in a careful study of the domains of operators. 
% Intuitively, when we say that $H$ is a 
% hermitian (and self-adjoint) operator, it really means that it is so when $H$ acts on functions $\varphi(x)$ 
% which ``behave well'' in a certain sense. In our case, since $H$ is essentially a second-derivative operator,
% the ``well behaving'' functions are those for which $\varphi''(x)$ well behaves everywhere.
% But our $\psi(x)$, obeying (\ref{psi_chi}), is {\em not} such a function, because $\psi''(x)$
% does not behave well at the boundary of the interval $D=(a,b)$. 
% More precisely, in general, if ${\cal D}(H)$ is the domain of the operator $H$ on which it is hermitian and 
% self-adjoint, and if $\varphi\in {\cal D}(H)$, then $\chi_{\bar{D}} \varphi$ does not necessarily need to obey 
% $\chi_{\bar{D}} \varphi\in {\cal D}(H)$. Instead, we can have
% \begin{equation}\label{notin}
%  \chi_{\bar{D}} \varphi\notin {\cal D}(H). 
% \end{equation}
% Indeed, the wave function obeying (\ref{psi_chi}) can be written in the form 
% \begin{equation}
%  \psi(x)=\chi_{\bar{D}}(x) \varphi(x) ,
% \end{equation}
% which is why it obeys the pathological property (\ref{notin}). But it does not mean that such a wave function 
% does not make mathematical or physical sense. It only means that, for such a wave function, the operator $H$ 
% is not hermitian. As we shall see, such a wave function may have a sensible physical interpretation.

\section{Avoiding quantum Zeno effect}
\label{SECzeno}
 
We have seen in Sec.~\ref{SECexp} that, under certain conditions, the average value of the hermitian Hamiltonian
can have a non-zero imaginary part. As a consequence, one might expect that such a quantum system should behave 
even less classically than typical quantum systems.
Surprisingly, however, we shall see that such an imaginary part can actually imply a {\em more} classical behavior.

\subsection{A basic demonstration}
\label{SECzenobasic}

Consider a quantum system initially prepared in the state $|\psi_0\rangle$, then left to evolve with the Hamiltonian $H$ during 
the time $\delta t$, and finally measured. What is the probability that upon measurement it will again be found in the state 
$|\psi_0\rangle$? For arbitrary $\delta t$, this probability is
\begin{equation}
p_0=\left|c_0(\delta t)\right|^2 ,
\end{equation}
where
\begin{equation}
c_0(\delta t)=\langle\psi_0|e^{-iH\delta t}|\psi_0\rangle .
\end{equation}
%(Here, as in the rest of the paper, we work in units $\hbar=1$.) 
But we are interested in the case of small $\delta t$, so the expansion in $\delta t$ gives
\begin{equation}\label{P0exp}
p_0=1+i\delta t [\langle\psi_0|H|\psi_0\rangle^* -\langle\psi_0|H|\psi_0\rangle] +{\cal O}(\delta t^2) .
\end{equation}
Normally one would expect that $\langle\psi_0|H|\psi_0\rangle$ is real so that the square bracket in
(\ref{P0exp}) is zero, implying $p_0=1+{\cal O}(\delta t^2)$.
The absence of the term linear in $\delta t$ is well known to imply the quantum Zeno effect 
\cite{misra,home-whitaker,decoh1,zeno-review,auletta}: 
If measurements are performed infinitely often (that is, the time between measurements $\delta t\to 0$), 
then the system stays in the initial state $|\psi_0\rangle$ forever. After an arbitrarily long time,
the probability that the system will be found in any state orthogonal to $|\psi_0\rangle$ vanishes.

However, we have seen in Sec.~\ref{SECexp} that, under certain conditions,
$\langle\psi_0|H|\psi_0\rangle$ does {\em not} need to be real. When these conditions are met, we can have
\begin{equation}\label{averim}
 \langle\psi_0|H|\psi_0\rangle=E_0-i\frac{\gamma}{2} ,
\end{equation}
where $E_0$ and $\gamma$ are real parameters. Consequently, (\ref{P0exp}) becomes
\begin{equation}\label{P0exp2}
p_0=1 -\gamma \delta t +{\cal O}(\delta t^2) .
\end{equation}
To understand the physical meaning of this, we consider $k$ repeated measurements performed during the time $t=k\delta t$.
In the limit $\delta t\to 0$, $k\to\infty$, with $t=k\delta t$ finite, the joint probability that the system will 
each time be found in the state $|\psi_0\rangle$ is
\begin{equation}\label{p0k}     
\lim_{k\to\infty} p_0^k=\lim_{k\to\infty} \left( 1-\gamma \frac{t}{k} \right)^k = e^{-\gamma t} .
\end{equation}
Thus, despite the limit $\delta t \to 0$,
there is a nonzero probability $1-e^{-\gamma t}$ that, after a finite time $t$, the system will be found in 
a state orthogonal to $|\psi_0\rangle$.
This shows that, when (\ref{averim}) has a non-zero imaginary part, 
{\em the quantum Zeno effect can be avoided}.

Finally, note that (\ref{P0exp2}) and (\ref{p0k}) only make sense in the regime in which probability 
is a number in the interval $[0,1]$. Specifically, if $\gamma>0$, then (\ref{p0k}) only makes sense 
for $t\geq 0$, while (\ref{p0k}) for $t<0$ does not describe any physical process. 
In general, understanding physics at times for which a formula for probability gives a value 
out of the interval $[0,1]$ requires a physical insight beyond that formula. 
This will have an important consequence in the study of arrival time in Sec.~\ref{SECarrival}. 

\subsection{More general formulation}\label{SECmore}

Now we want to understand the avoidance of the quantum Zeno effect in terms of the general formalism 
developed in \cite{nik_tajron}.
  
In an arbitrary quantum system, consider a projective measurement with only two possible outcomes; 
one possible outcome corresponds to some projector $\pi$ and the other to the projector $\bar{\pi}=1-\pi$.
The measurement is repeated many times, while between measurements the system evolves with the Hamiltonian $H$.
We want to find the joint probability that, at $k$ subsequent measurements
at times $t_1=\delta t$, $t_2=2\delta t$, ..., $t_k=k\delta t$, 
the outcome will each time turn out to be
the same, say the one corresponding to $\bar{\pi}$. This probability is given by the 
compact formula \cite{nik_tajron}
\begin{equation}\label{probk}
 \bar{P}_k = ||V^k|\psi_0\rangle||^2 ,
\end{equation}
where $|\psi_0\rangle$ is the initial state at $t=0$ and
\begin{equation}\label{defV}
 V \equiv \bar{\pi}e^{-iH\delta t} .
\end{equation}
The goal is to find the operator $V^k$ in the limit $\delta t\to 0$, $k\to\infty$, with $t=k\delta t$ finite.
First we write
\begin{eqnarray}\label{Vder1}
 V^{k+1} &=& VV^k=\bar{\pi}e^{-iH\delta t}V^k
\nonumber \\
&=& \bar{\pi} \left[ 1-iH\delta t + {\cal O}(\delta t^2) \right] V^k ,
\end{eqnarray}
and ignore ${\cal O}(\delta t^2)$ since we are interested in the $\delta t\to 0$ limit. 
Then we use $\bar{\pi}V^k=V^k$ to write (\ref{Vder1}) as
\begin{equation}
 V^{k+1}=V^k-i\delta t \, \bar{\pi}H\bar{\pi} V^k ,
\end{equation}
which can be rewritten as
\begin{equation}\label{Vder3}
 \frac{V^{k+1}-V^k}{\delta t}=-i\overline{H} V^k ,
\end{equation}
where $\overline{H}$ is defined as in (\ref{projected}). Finally, using the notation 
$V^k\equiv V(k\delta t)=V(t)$, we see that (\ref{Vder3}) in the $\delta t \to 0$ limit
is the differential equation
\begin{equation}
 \frac{dV(t)}{dt}=-i\overline{H}V(t) ,
\end{equation}
with the solution
\begin{equation}
 V(t)=e^{-i\overline{H}t}V(0)=e^{-i\overline{H}t}\bar{\pi}.
\end{equation}
Thus the final result can be written as 
\begin{equation}\label{Vk}
 \lim_{\delta t\to 0}V^k=e^{-i\overline{H}t}\bar{\pi} ,
\end{equation}
where $k=t/\delta t$. The derivation of (\ref{Vk}) above is a somewhat simpler version 
of the derivation presented in \cite{nik_tajron}.

Now let us apply this result to the quantum Zeno effect. In finite dimensional Hilbert spaces
the projected Hamiltonian $\overline{H}$ is always hermitian, which can be seen from 
\begin{equation}\label{herm}
\overline{H}^\dagger= (\bar{\pi}H\bar{\pi})^{\dagger}
=\bar{\pi}^{\dagger}H^{\dagger}\bar{\pi}^{\dagger}=\bar{\pi}H\bar{\pi} =\overline{H} .
\end{equation}
In infinite dimensional Hilbert spaces (\ref{herm}) is not always correct, but in most cases 
of physical interest it is correct. Hence, in most cases of physical interest,  
$\overline{H}$ is hermitian. In such cases $e^{-i\overline{H}t}$ is unitary, 
so (\ref{probk}) with (\ref{Vk}) gives
\begin{equation}\label{probk2}
 \lim_{\delta t\to 0} \bar{P}_k = ||\bar{\pi}|\psi_0\rangle||^2 ,
\end{equation}
which is {\em time independent}. In particular, if the initial state obeys $\bar{\pi}|\psi_0\rangle=|\psi_0\rangle$,
then (\ref{probk2}) reduces to
\begin{equation}\label{probk3}
 \lim_{\delta t\to 0} \bar{P}_k = 1 ,
\end{equation}
which is the quantum Zeno effect. For any finite time $t=k\delta t$, the system remains in the initial state with certainty.
The infinitely frequent measurements prevent any change. 

This conclusion, however, can be avoided if the projected Hamiltonian $\overline{H}$ is not hermitian, because then 
$e^{-i\overline{H}t}$ is not unitary so (\ref{probk2}) is no longer true.
Indeed, we have already seen in Sec.~\ref{SECproj} that non-hermitian $\overline{H}$ leads to an imaginary part  
of the expectation value of $H$, which leads to an avoidance of quantum Zeno effect as demonstrated in Sec.~\ref{SECzenobasic}.
Here we want to understand the avoidance of quantum Zeno effect from a more general point of view.
For that purpose, in the following we use some probability theory that is equally valid in both classical and quantum physics.

First, let us introduce some language that treats classical and quantum measurement outcomes on an equal footing: 
the measurement outcome associated with $\pi$ will be called {\em positive outcome},
while the measurement outcome associated with $\bar{\pi}$ will be called {\em negative outcome}.
The probability $\bar{P}_k$ on the left-hand side of (\ref{probk}) can be written as \cite{nik_tajron}
\begin{equation}\label{Pkprod}
 \bar{P}_k = \prod_{j=1}^k \bar{p}(t_j|t_{j-1}) ,
\end{equation}
where $\bar{p}(t_j|t_{j-1})$ is the conditional probability of negative outcome at time $t_j$, 
given that the outcome was negative at time $t_{j-1}$. It is convenient to write this in the exponential form
\begin{equation}
 \bar{P}_k = \exp\left( \sum_{j=1}^k \ln \bar{p}(t_j|t_{j-1}) \right),
\end{equation}
so in the limit $\delta t \to 0$ this becomes
\begin{equation}\label{barPk}
 \lim_{\delta t\to 0} \bar{P}_k = e^{-\int_0^t dt'w(t')} ,
\end{equation}
where
\begin{equation}\label{w(t)}
 w(t)\equiv - \lim_{\delta t\to 0} \frac{\ln \bar{p}(t|t-\delta t)}{\delta t} .
\end{equation}
Next we use $\bar{p}(t|t-\delta t)=1-p(t|t-\delta t)$, where $p(t|t-\delta t)$ is the conditional probability 
of {\em positive} outcome at time $t$, given that the outcome was negative at time $t-\delta t$.
Clearly, $p(t|t)=0$, so $p(t|t-\delta t)\ll 1$ for small $\delta t$ and we have
$\ln \bar{p}(t|t-\delta t) = \ln [1-p(t|t-\delta t)] \simeq -p(t|t-\delta t)$. Thus (\ref{w(t)}) can be written as 
\begin{equation}\label{w(t)2}
 w(t) =\lim_{\delta t\to 0} \frac{p(t|t-\delta t)}{\delta t} .
\end{equation} 
We see that $w(t)$ is not negative, provided that the probability $p(t|t-\delta t)$
is not negative. 

Now we can distinguish two cases. 
The first case is $w(t)=0$ for all $t$, in which case (\ref{barPk}) reduces to
\begin{equation}\label{barPk2}
 \lim_{\delta t\to 0} \bar{P}_k = 1 .
\end{equation}
In this case, from (\ref{w(t)2}) we see that
$p(t|t-\delta t)$ has expansion in $\delta t$ of the form 
\begin{equation}
 p(t|t-\delta t)= f(t) (\delta t)^2 + \ldots 
\end{equation}
for some function $f(t)$, 
without having a term linear in $\delta t$.
In the quantum context, (\ref{barPk2}) corresponds to the quantum Zeno effect.
Note that even a negative outcome can induce a quantum Zeno effect \cite{home-whitaker,zeno-review}.

The second case is $w(t)>0$ for some $t$, in which case (\ref{barPk}) implies
\begin{equation}\label{barPk3}
 \lim_{\delta t\to 0} \bar{P}_k < 1 .
\end{equation}     
In this case, from (\ref{w(t)2}) we see that $p(t|t-\delta t)$
has expansion in $\delta t$ of the form 
\begin{equation}
 p(t|t-\delta t)=w(t)\,\delta t + \ldots .
\end{equation}
In the quantum context, (\ref{barPk3}) corresponds to {\em avoidance of the quantum Zeno effect}. 
It is a generalization of the example worked out in Sec.~\ref{SECzenobasic},
where $w(t) \equiv \gamma$ is a constant.

\subsection{Application to arrival time}

The analysis so far shows that avoiding quantum Zeno effect is possible mathematically. 
Now we want to discuss a physical realization of this mathematical possibility.
We shall show that quantum arrival time as described in \cite{nik_tajron} 
can be understood as a physical system where the the quantum Zeno effect is avoided.

Consider a wave packet impinging from the left towards the detector at $x=0$. At each time $t_j=j\delta t$, $j=1,2,\ldots$,
the detector checks whether the particle has arrived. If it has arrived then the detector clicks, which corresponds to the positive 
measurement outcome associated with a projector $\pi$ \cite{jur-nik,nik_tajron}. If it has not arrived 
then the detector does not click, which corresponds to the negative 
measurement outcome associated with a projector $\bar{\pi}=1-\pi$. If the particle has not arrived then we know that it is still 
in the region $x<0$ on the left from the detector, so the projector $\bar{\pi}$ is given by (\ref{barpi})
with $\bar{D}=(-\infty,0)$. Now consider the limit $\delta t\to 0$. If the quantum Zeno effect was present, 
it would mean that the particle would stay in the region $\bar{D}=(-\infty,0)$ forever, and never arrive at the detector at $x=0$.
Yet it was shown in \cite{nik_tajron} that there is a non-zero probability of 
arrival in the limit $\delta t \to 0$. The results of this paper reveal that this possibility of arrival is precisely
an example of avoidance of the quantum Zeno effect.  

A full analysis of the arrival time probability, much beyond the results obtained in \cite{jur-nik,nik_tajron},
is presented in Sec.~\ref{SECarrival}.

\section{Gambler's fallacy and passive quantum measurement}
\label{SECgambler}

In Sec.~\ref{SECzeno} we have discussed how quantum measurement can behave like a classical measurement, in the
context of avoidance of the quantum Zeno effect in the $\delta t\to 0$ limit.
However, to understand the classical behavior of such a quantum measurement, the limit $\delta t\to 0$ is not essential.
To further elucidate what is going on, in this section we discuss phenomena that do not depend on that limit.
For the sake of intuition, we frame the discussion in the context of gambler's fallacy.  

\subsection{Classical gambler's fallacy}

The classical gambler's fallacy, or simply gambler's fallacy (see e.g. \cite{tijms}), is an elementary 
example of erroneous reasoning in probability theory. Suppose, for instance, that a gambler
playing roulette observed that several times in a row the outcome was an even number.
Since even and odd numbers should, in average, appear equally often, the gambler expects 
that now the probability of an odd number is larger than $1/2$. More generally, if a certain random 
outcome have not appeared for a long time, the gambler expects that now the probability of that outcome 
is larger than it was at the beginning. 

In the case of roulette such an expectation is clearly wrong, this is an example of gambler's fallacy.
However, such a reasoning is not always wrong. For instance, if someone is waiting for the bus and the bus
has not arrived for a long time, it is justified to expect that the bus should arrive soon. 
The longer one waits for the bus, the larger is the time probability density of its arrival. This is an example
where the gambler's reasoning is correct, i.e. where the gambler's fallacy is avoided.

Where does the difference between roulette and bus come from? 
Each time after the gambler sees the roulette outcome, the roulette is {\em reset}  
to the initial state in which every number is {\it a priori} equally probable.
The bus, by contrast, is not reset. When we observe that the bus has not arrived yet, 
the bus is not suddenly moved to its initial station. Our observations have no influence on the 
bus motion at all. Each time when we observe that the bus has not yet arrived we update our {\em knowledge} 
on the bus, but it has no influence whatsoever on the bus {\em itself}.

\subsection{Quantum gambler's fallacy}
\label{SECqgambler}

The roulette case and the bus case have also their quantum analogies. 

A quantum analogy of the roulette case is a decay of an unstable atom.
To a good approximation, the decay obeys an exponential law according to which the probability that 
the atom will not decay during the time $\Delta t$ is equal to $e^{-\Gamma \Delta t}$, where $\Gamma$ is a constant.
But if the atom has not decayed for a long time $\Delta t\gg\Gamma^{-1}$, the probability of its decay is not any 
larger. More precisely, if one observed that the atom has not yet decayed at time $t_1$,
the probability that it will not decay in the future of $t_1$,
between times $t_2$ and $t_1$, is $e^{-\Gamma (t_2-t_1)}$. 
Someone who would expect that the probability of a decay has increased because the decay has not happened 
for a long time would commit a fallacy, very much similar to the classical gambler's fallacy. 
We refer to fallacy of that type as a quantum gambler's fallacy. 

A quantum analogy of the bus case is arrival of a quantum particle.
Clearly, just as for the bus, if the particle has not arrived for a long time, the probability 
density of arrival is larger than it was at the beginning. 
Thus, for the quantum arrival, the gambler's reasoning is correct, i.e. the quantum gambler's fallacy is avoided.  

But were does the difference between quantum decay and quantum arrival come from, exactly?
The origin of the difference is completely analogous to that in their classical 
cousins, which we explain next.

\subsection{How can quantum measurement behave like a classical measurement?}

Consider first the quantum decay. If the initial state of an unstable atom is $|\psi_0\rangle$, then,
each time when we observe that the particle has not decayed yet, the projective measurement projects the state
with the projector $\bar{\pi}=|\psi_0\rangle\langle\psi_0|$, which resets the system back to its initial state 
$|\psi_0\rangle$. The quantum measurement is associated with a {\em physical} change, which is analogous to the fact 
that the state of roulette is physically changed each time after the roulette outcome.
Both the unstable atom and the roulette are reset to its initial state.
Indeed, it is a general feature of quantum measurements that they induce a physical change. 
In this sense quantum measurement is very different from a classical measurement.

But how then the quantum arrival can be analogous to the bus arrival? 
How can it be that a quantum measurement can be interpreted as a mere update of knowledge,
without producing a physical change? In other words, how can a quantum measurement 
behave like a classical measurement? The answer lies in a very special property of the projector (\ref{barpi}) 
represented by the characteristic function (\ref{pc2}). 
The distributional derivatives $\chi'_{\bar{D}}(x)$ and $\chi''_{\bar{D}}(x)$ vanish 
everywhere except at the boundary of $\bar{D}$. 
On the other hand, since $\bar{D}$ is open, $\chi_{\bar{D}}(x)$ vanishes at the boundary of ${\bar{D}}$.
Hence $\chi_{\bar{D}}(x)\chi'_{\bar{D}}(x)$ and $\chi_{\bar{D}}(x)\chi''_{\bar{D}}(x)$ vanish everywhere, so 
for any well behaved wave function $\varphi(x)$ we have \cite{nik_tajron}
\begin{eqnarray}\label{piHpi}
 \bar{\pi}H\bar{\pi}\varphi(x) &=& \chi_{\bar{D}}(x)\frac{-1}{2m}\frac{d^2}{dx^2}(\chi_{\bar{D}}(x)\varphi(x))
\nonumber \\
 &=& \chi_{\bar{D}}(x)\frac{-1}{2m}\frac{d^2\varphi(x)}{dx^2} 
\nonumber \\
 &=& \bar{\pi}H\varphi(x) ,
\end{eqnarray}
where in the second line we used $\chi_{\bar{D}}(x)\chi'_{\bar{D}}(x)=0$ and $\chi_{\bar{D}}(x)\chi''_{\bar{D}}(x)=0$.
Since (\ref{piHpi}) does not depend on $\varphi(x) \in\mathcal{D}(H)$, we get a surprising formula
\begin{equation}\label{magic}
 \bar{\pi}H\bar{\pi}=\bar{\pi}H ,
\end{equation}
or equivalently
\begin{equation}\label{magic_com}
 \bar{\pi}[H,\bar{\pi}]=0 .
\end{equation}
This is surprising because $H$ and $\bar{\pi}$ do {\em not} commute
\begin{equation}\label{nocom}
[H,\bar{\pi}]\neq 0 .
\end{equation}
If $H$ and $\bar{\pi}$ were treated like hermitian matrices in a finite dimensional Hilbert space,
then hermitian conjugation of (\ref{magic}) would imply $\bar{\pi}H\bar{\pi}=H\bar{\pi}$, 
which together with (\ref{magic}) would imply $H\bar{\pi}=\bar{\pi}H$, i.e. $[H,\bar{\pi}]= 0$,
so (\ref{magic}) would be incompatible with (\ref{nocom}).
Thus (\ref{magic}) is surprising because, under the condition that $H$ and $\bar{\pi}$ are non-commuting hermitian 
operators, (\ref{magic}) is only possible in the infinite dimensional Hilbert space
\footnote{Actually, the commutator $[H, \bar{\pi}]$ is ill-defined and one should work only with the Weyl 
form of the operators when considering commutators. In this case the Weyl form for the Hamiltonian 
is the unitary operator $U(t)=e^{-iHt}$, which is bounded and therefore the commutation relations make sense 
for all elements of the Hilbert space. Thus the commutators are only rigorously defined among the
following operators: the 
projectors $\pi$ and $\bar{\pi}$, the unitary operator $U(t)$ and 
the contraction operator $V(t)$.}.

The surprising property (\ref{magic}) also gives a hint why a measurement defined by the projector 
$\bar{\pi}$ behaves like a classical measurement. Heuristically, if $\bar{\pi}$ was a classical commuting quantity 
we could write $\bar{\pi}H\bar{\pi}=\bar{\pi}\bar{\pi}H=\bar{\pi}H$, which could be thought of as a heuristic 
``derivation'' of (\ref{magic}). 

The classical behavior of the measurement defined by the projector $\bar{\pi}$ is shown explicitly by 
the following analysis. It was shown in \cite{nik_tajron} that (\ref{magic}) implies
\begin{equation}\label{magic2}
\left( \bar{\pi}e^{-iH\delta t} \right)^k= \bar{\pi}e^{-iHt} ,
\end{equation}
where $t=k\delta t$, so (\ref{defV}) implies
\begin{equation}\label{magic2.1}
V^k= \bar{\pi}e^{-iHt} .
\end{equation}
We want to find how the quantum state is changed by $k$ repeated measurements, 
under the condition that each time the measurement outcome turned out to be negative, 
namely described by the projector $\bar{\pi}$. Under this condition, the state after $k$ measurements is the conditional state 
\begin{equation}\label{psi_c}
 |\bar{\psi}_c(t_k)\rangle = \frac{V^k|\psi_0\rangle}{||V^k|\psi_0\rangle||}
= \frac{\bar{\pi}e^{-iHt}|\psi_0\rangle}{||\bar{\pi}e^{-iHt}|\psi_0||} .
\end{equation}
But this can be written as 
\begin{equation}\label{psi_c2}
 |\bar{\psi}_c(t)\rangle =\frac{\bar{\pi}|\psi(t)\rangle}{||\bar{\pi}|\psi(t)\rangle||} ,
\end{equation}
where $t=k\delta t$ and 
\begin{equation}\label{psinomeas}
 |\psi(t)\rangle=e^{-iHt}|\psi_0\rangle
\end{equation}
is the state in the {\em absence} of measurement. Hence the state (\ref{psi_c2}) cannot be distinguished
from a scenario in which the following happened: the system evolved {\em without} measurements at all times
$t_1,\ldots,t_{k-1}$, and only at the final time $t=t_k$ a measurement has been performed 
with the negative outcome defined by $\bar{\pi}$. In other words, {\em at any time $t_k$, the state looks 
as if there were no any measurements at earlier times $t_1,\ldots t_{k-1}$}. 
This means that earlier measurements did not really affect the state, 
which makes any such measurement 
with a negative outcome very similar to a classical passive measurement. 
For that reason, such a measurement 
with negative outcome is called {\em passive quantum measurement}. 

We emphasize that manifestation of a passive quantum measurement is very different 
from manifestation of an ordinary quantum measurement. While ordinary quantum measurement usually affects the state,
a passive quantum measurement never affects the state. In particular, an ordinary quantum 
measurement can reset the system to the initial state, but a passive quantum measurement cannot do that.
This explains how the quantum arrival can be analogous to the bus arrival;
the measurements before the quantum arrival are passive quantum measurements, so they
do not reset the system to the initial state.

Note that (\ref{piHpi}) is derived for a free Hamiltonian of one particle in one dimension,
but a completely analogous derivation is valid much more generally, e.g. for a Hamiltonian
of the form 
\begin{equation}
H=\frac{{\bf p}^2}{2m}+V({\bf x})
\end{equation}
describing a particle in 3 dimensions with arbitrary interaction $V({\bf x})$. 
The only essential assumptions are (i) that the momenta are represented by derivative
operators and (ii) that $\bar{D}$ is an open set, in the 
configuration space of particle positions.

Finally note that in the analysis above, by the absence of measurement we do not mean the absence of the detector. The presence of the detector is encoded in the Hamiltonian $H$. The presence of the 
measurement means the presence of the projection with $\bar{\pi}$. Clearly, it is possible 
to consider the unitary evolution governed by $H$ without considering the non-unitary projection. 
Hence, when we say that the system behaves as if the measurement was absent, 
it does not mean that the system behaves as if the detector was absent.
For example, if we place a detector at one of the slits in a double-slit experiment,
then the presence of the detector will destroy the interference, which depends 
on the Hamiltonian describing the detector, 
but not on the presence or absence of the projection by $\bar{\pi}$.

\subsection{Probabilistic aspects}

Now let us discuss passive quantum measurements from a probabilistic point of view.
Eqs.~(\ref{magic2.1}) and (\ref{psinomeas}) imply that (\ref{probk}) can be written as
\begin{equation}\label{probkt}
 \bar{P}(t) = ||\bar{\pi}e^{-iHt}|\psi_0\rangle||^2 = ||\bar{\pi}|\psi(t)\rangle||^2.
\end{equation}
Using 
\begin{equation}
\langle x|\bar{\pi}|\psi(t)\rangle = \chi_{\bar{D}}(x)\psi(x,t) ,
\end{equation}
where $\psi(x,t)=\langle x |\psi(t)\rangle$ is the wave function in the absence of measurement,
(\ref{probkt}) can be written as
\begin{eqnarray}\label{probkt2}
 \bar{P}(t) &=& \int_{-\infty}^{\infty} dx\, \chi_{\bar{D}}(x)\, \psi^*(x,t) \psi(x,t)
\nonumber \\
&=& \int_{\bar{D}} dx\, |\psi(x,t)|^2 .
\end{eqnarray}
In the context of quantum arrival time, ${\bar{D}}$ is the open region $\bar{D}=(-\infty,0)$
on the left from the detector and (\ref{probk}) is the probability
that the particle has not yet arrived at the time $t=k\delta t$.
Eq.~(\ref{probkt2}) shows that this probability is defined by the probability density $|\psi(x,t)|^2$ 
in the {\em absence} of measurement, demonstrating again that the repeated projections by $\bar{\pi}$, appearing in 
(\ref{probk}) with (\ref{defV}), did {\em not} affect the wave function $\psi(x,t)$.

% For completeness, let us also write a more explicit expression for the numerator in (\ref{cond_final2}).
% In general, it is 
% \begin{equation}\label{Pkfinal}
%  P_k=||\pi e^{-iH\delta t} V^{k-1}|\psi_0\rangle||^2 .
% \end{equation}
% Eq.~(\ref{magic2.1}) implies
% \begin{equation}
%  V^{k-1}|\psi_0\rangle= \bar{\pi}e^{-iH(k-1)\delta t}|\psi_0\rangle = \bar{\pi} |\psi(t-\delta t)\rangle ,
% \end{equation}
% so (\ref{Pkfinal}) can be written as
% \begin{equation}\label{Pkfinal2}
%  P_k=||\pi e^{-iH\delta t} \bar{\pi} |\psi(t-\delta t)\rangle ||^2 .
% \end{equation}
% The measurement associated with $\pi$, corresponding to a detection of the particle, is not passive. 
% Therefore, in general, a further simplification of (\ref{Pkfinal2}) is not an easy job. 
% In some cases, a significant simlification of (\ref{Pkfinal2}) is possible 
% in an approximative sense \cite{nik_tajron,jur-nik}.

\section{Arrival time distribution}
\label{SECarrival}

\begin{figure}[t]
%\centering
\includegraphics[width=7cm]{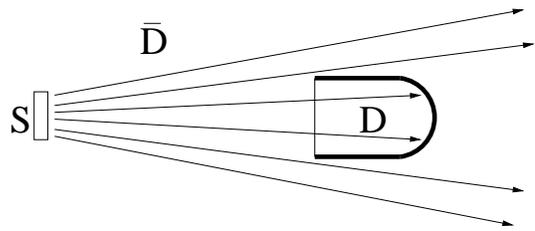}
\caption{\label{fig1}
A semi-realistic example of an arrival experiment.
The small `\textsf{D}'-shaped detector region $D$ is surrounded by its complement ${\bar D}$. 
The detector detects a particle arriving at $D$ from the source $S$.
The boundary $\partial\bar{D}$ of $\bar{D}$ is the same as the boundary $\partial D$ of $D$.
The bold $\mbox{\boldmath `$\supset$}$'-shaped part of the boundary absorbs particles from the inside.
The lines with arrows represent the stream of particles. For classical particles they are straight lines,
so the flux through the boundary from the outside comes only from the front (left) `$|$'-shaped part of the boundary,
where the flux is positive. For quantum particles the ``stream lines'', namely integral curves of the probability current,
may wiggle due to interference (not shown in the picture), 
but in most cases of practical interest the wiggling is not too extreme, so that 
the total flux through the boundary from the outside is still positive.
}
\end{figure}

Now we study the problem of arrival time distribution in detail.
This problem has a long history, see e.g. \cite{muga-physrep,V1.10,V2.4,maccone} for reviews.
Here we shall first propose the solution of the problem using only elementary physics,
after which we shall show that the same result is obtained with our theory of passive
quantum measurements. An example of an arrival experiment that one can have in mind 
is depicted in Fig.~\ref{fig1}.

\subsection{Arrival and departure time distribution from elementary physics}
\label{SECsimple}

Let us start with some probabilistic insights which
are equally valid in classical and quantum physics.
Consider a connected region $\bar{D}$ in 3-dimensional space, with a 2-dimensional boundary $\partial \bar{D}$ 
(for an example, see Fig.~\ref{fig1}).
The region $\bar{D}$ is defined as an open set, in the sense that it does not contain its boundary $\partial \bar{D}$.
Let $\bar{P}(t)$ be the probability that, at time $t$, the particle is in $\bar{D}$, and let $P(t)=1-\bar{P}(t)$
be the probability that the same particle at the same time is in $D$ defined as the complement of $\bar{D}$.
In general, the two probabilities change with time so that
\begin{equation}
 \frac{d\bar{P}(t)}{dt}=-\frac{dP(t)}{dt} .
\end{equation}
Thus, unless both $d\bar{P}/dt$ and $dP/dt$ are zero, we have that one of those derivatives is positive and the other negative.

Now let us apply some quantum way of reasoning in an implicit form.
The case $dP/dt>0$ means that the probability of being in $D$ {\em increases}, 
so we can say that the particle {\em arrives} to $D$.
Conversely, the case $dP/dt<0$ means that the probability of being in $D$ {\em decreases},
so we can say that the particle {\em departs} from $D$. 
This motivates us to define the arrival probability density ${\cal P}_{\rm arr}$ 
and the departure probability density ${\cal P}_{\rm dep}$
by the following formulas
\begin{eqnarray}\label{arr_dep0}
 {\cal P}_{\rm arr}(t)=\frac{dP(t)}{dt}=-\frac{d\bar{P}(t)}{dt}, \;\;\; {\rm when} \;\; \frac{dP(t)}{dt}>0 ,
\nonumber \\
 {\cal P}_{\rm dep}(t)=\frac{d\bar{P}(t)}{dt}=-\frac{dP(t)}{dt}, \;\;\; {\rm when} \;\; \frac{d\bar{P}(t)}{dt}>0 .
\end{eqnarray}
Another way to write this is
\begin{equation}\label{arr_dep}
 \frac{d\bar{P}(t)}{dt}=\left\{ 
\begin{array}{rl}
 \displaystyle -{\cal P}_{\rm arr}(t) & \;\; {\rm for} \;\;  d\bar{P}/dt<0 \\
 {\cal P}_{\rm dep}(t) & \;\; {\rm for} \;\; d\bar{P}/dt>0 . 
\end{array}
\right.
\end{equation}
In words, negative $d\bar{P}/dt$ physically represents arrival probability density (times $-1$), while 
positive $d\bar{P}/dt$ physically represents departure probability density.  

What is quantum about the reasoning above? To understand this, consider the two straight lines 
that enter $D$ in Fig.~\ref{fig1} and imagine that one of them, say the lower one, has an arrow 
with the opposite orientation than actually shown in Fig.~\ref{fig1}. 
In classical physics it would mean that the upper 
particle arrives to $D$, while the lower particle departs from $D$, which demonstrates
a possibility that both arrival and departure probability are non-zero at the same time.
Hence (\ref{arr_dep0}) and (\ref{arr_dep}), which say that arrival and departure probability
cannot both be non-zero, are in contradiction with classical physics. 
In the quantum case, however, the detector in Fig.~\ref{fig1} only determines 
whether the particle is in $D$ or not, it cannot distinguish the two straight lines that 
enter $D$, which is why (\ref{arr_dep0}) and (\ref{arr_dep}) are compatible 
with quantum mechanics. In the standard interpretation of quantum mechanics 
one would say that particles do not have trajectories, so
the two straight lines do not correspond to anything physical.
In the Bohmian interpretation 
(see e.g. \cite{nik-ibm} and references therein)
one would say that the two straight lines 
correspond to actual trajectories, which however are not measured 
by the detector in Fig.~\ref{fig1}. The two interpretations tell us different stories,
but they 
make the same measurable predictions on {\em detected} arrivals \cite{jur-nik}.
Thus, as long as (\ref{arr_dep0}) and (\ref{arr_dep}) are viewed  
strictly operationally as probabilities of {\em detection}, 
their validity does not depend on which interpretation of quantum mechanics
one uses.

Now let us introduce some elementary quantum mechanics
in an explicit form.
Suppose that at the initial time $t=0$ the particle is inside the region $\bar{D}$, i.e. that the initial wave function 
$\psi({\bf x},0)$ vanishes for all ${\bf x}\in D$. 
Thus the normalization of $\psi$ at the initial time gives
\begin{equation}\label{norm}
 \int_{\bar{D}} d^3x\, |\psi({\bf x},0)|^2 =1 .
\end{equation}
Here $|\psi({\bf x},t)|^2\equiv \rho({\bf x},t)$ is the standard quantum probability density, satisfying the usual 
continuity equation
\begin{equation}\label{cont}
 \frac{\partial \rho({\bf x},t)}{\partial t} + \mbox{\boldmath $\nabla$}\cdot {\bf j}({\bf x},t) =0,
\end{equation}
where ${\bf j}({\bf x},t)$ is the usual quantum probability current.
At time $t$, 
the probability $\bar{P}(t)$ that the particle is in $\bar{D}$ is given by
\begin{equation}\label{barP(t)}
 \bar{P}(t)=\int_{\bar{D}} d^3x\, |\psi({\bf x},t)|^2 .
\end{equation} 
Essentially, (\ref{norm}), (\ref{cont}) and (\ref{barP(t)}) is all what we need of quantum mechanics
in this subsection.
Thus (\ref{barP(t)}) and (\ref{cont}) give
\begin{eqnarray}\label{FINAL}
 - \frac{d\bar{P}(t)}{dt} &=& -\int_{\bar{D}} d^3x\, \frac{\partial\rho({\bf x},t)}{\partial t} = 
\int_{\bar{D}} d^3x\, \mbox{\boldmath $\nabla$} \cdot {\bf j}({\bf x},t) 
\nonumber \\
&=& \int_{\partial \bar{D}} d{\bf S}\cdot {\bf j}({\bf x},t) 
\equiv \Phi_{\partial \bar{D}}(t),
\end{eqnarray}
where $d{\bf S}$ is the area element directed {\em outwards} from $\bar{D}$, and 
$\Phi_{\partial \bar{D}}(t)$ is the flux of the probability current ${\bf j}$ through the boundary $\partial \bar{D}$ 
at time $t$. 

Now we can combine (\ref{FINAL}) with (\ref{arr_dep}) to find the formula for the arrival time distribution.
When $\Phi_{\partial \bar{D}}(t)>0$, then ${\cal P}_{\rm arr}(t)=\Phi_{\partial \bar{D}}(t)$.
When $\Phi_{\partial \bar{D}}(t)<0$, then there is no arrival at all 
(because the particle exhibits departure, rather than arrival), so ${\cal P}_{\rm arr}(t)=0$. All together,
the final compact formula for the arrival time distribution is
\begin{equation}\label{Parr_fin}
 {\cal P}_{\rm arr}(t)=\left\{ 
\begin{array}{cl}
 \displaystyle \Phi_{\partial \bar{D}}(t) & \;\; {\rm for} \;\; \Phi_{\partial \bar{D}}(t)>0  \\
 0 & \;\; {\rm for} \;\; \Phi_{\partial \bar{D}}(t)\leq 0 . 
\end{array}
\right.
\end{equation}

The result that the arrival probability distribution can be related to the flux of the probability current is not new.
It has been obtained from standard quantum mechanics \cite{kijowski,delgado-muga,tumulka1}
and from Bohmian particle trajectories \cite{leavens,leavens_book,durrtime1,durrtime2}.
What is new in our analysis is the way we interpret the negative fluxes. In our analysis, 
negative flux corresponds to zero arrival probability density, which originates from the physical insight 
that in this case the particle departs, rather than arrives.  

\subsection{Consistent integration of arrival probability density}

Now let us study how the arrival probabilities can be consistently added, that is,
how the probability density ${\cal P}_{\rm arr}(t)$
is consistently integrated over time. Suppose that one wants to know  
the total probability that the particle will eventually arrive, at any time within the time interval $[0,T]$. 
If the flux is never negative, then from (\ref{FINAL}) we see that this probability is
\begin{eqnarray}\label{intprob0T}
 \int_0^{T} dt\, {\cal P}_{\rm arr}(t) &=& 
 -\int_{\bar{D}} d^3x \int_0^{T} dt\, \frac{\partial\rho({\bf x},t)}{\partial t}
\nonumber \\
&=& -\int_{\bar{D}} d^3x\, [\rho({\bf x},T)-\rho({\bf x},0)]
\nonumber \\
&=& 1-\int_{\bar{D}} d^3x\, \rho({\bf x},T) ,
\end{eqnarray}
where (\ref{norm}) has been used. Clearly, for any $T\geq 0$, this probability is a number in the interval $[0,1]$, 
as it should be. More generally, if the flux is never negative in the time 
interval $[T_1,T_2]$, then a generalization of (\ref{intprob0T}) implies that
\begin{eqnarray}\label{intprob}
\int_{T_1}^{T_2} dt\, {\cal P}_{\rm arr}(t) 
&=& \int_{\bar{D}} d^3x\, \rho({\bf x},T_1)-\int_{\bar{D}} d^3x\, \rho({\bf x},T_2)
\nonumber \\
&=& \bar{P}(T_1)-\bar{P}(T_2)
\end{eqnarray}
is the total probability that the particle will arrive within this interval.
Indeed, this integral is not negative (because ${\cal P}_{\rm arr}(t)$ is not negative)
and cannot exceed 1 (because the two integrals on the right-hand side are positive 
numbers that cannot exceed 1), so the probability is in the interval $[0,1]$.

But what if the flux is sometimes negative?
This is where the things become non-trivial and interesting.
For example, suppose that the flux is first positive in the time interval $(0,t_1)$, 
then negative in the interval $(t_1,t_2)$, and then positive again in the interval $(t_2,\infty)$.
Furthermore, suppose that $\int_0^{t_1} dt\,{\cal P}_{\rm arr}=1$, so that the particle arrives 
with certainty in the interval $(0,t_1)$. 
Then 
\begin{equation}
\int_0^{t_1} dt\,{\cal P}_{\rm arr}+\int_{t_2}^{T} dt\,{\cal P}_{\rm arr} >1 
\end{equation}
for $T>t_2$.
This, of course, does not mean that total probability can be larger than 1.
So what exactly does it mean? This question can be answered from two perspectives
depending on how one thinks of the arrival detector, but the two perspectives 
are observationally equivalent. Let us discuss the two perspectives separately. 

The first perspective is to think of detector as an idealized system 
that does not absorb the particle upon detection. 
This means that the same particle can arrive more than once, first in the interval
$(0,t_1)$, and then again in the interval $(t_2,T)$ after the departure during $(t_1,t_2)$.
The arrivals in the two intervals $(0,t_1)$ and $(t_2,T)$ are not mutually 
exclusive events, so their probabilities should not be added. The quantity ${\cal P}_{\rm arr}$ 
can be interpreted as arrival probability density in either of the intervals $(0,t_1)$ 
and $(t_2,T)$ separately, but not in their union $(0,t_1) \cup (t_2,T)$.
More generally, ${\cal P}_{\rm arr}$ is interpreted as arrival probability density in any 
interval $[T_1,T_2]$ during which the flux is never negative.      

The second, more realistic, perspective is to think 
of detector as a system that absorbs the particle upon detection.
But absorption during $(0,t_1)$ must be consistent with subsequent departure during $(t_1,t_2)$,
which means that the absorbed particle can be re-emitted during $(t_1,t_2)$. 
Effectively, as long as
one is only interested in the arrival probability density ${\cal P}_{\rm arr}$,
this makes the second perspective equivalent to the first.
The absorption and re-emission can  
be described by a second-quantization formalism in which absorption and re-emission 
are formally described as destruction and creation of a particle, respectively,
such that the total number of particles remains always conserved.
In such a description,
the particle that arrives in $(t_2,T)$ is viewed as a ``new'' particle, not as the ``same'' particle 
that arrived in $(0,t_1)$, but this formal difference between the two particles 
does not have observable consequences.
A detailed study of such a second-quantization formalism is beyond the scope 
of the present paper; it is planned for a future work.

\subsection{Arrival time distribution from passive quantum measurements}
\label{SECarrpass}

In Sec.~\ref{SECsimple} we have derived the arrival time probability density 
(\ref{Parr_fin}) without assuming any measurements 
(that is, projections). 
Now we shall see how the same result 
is obtained with passive quantum measurements.

We start from the study of the joint probability that the particle will eventually be detected at time $t_k$, 
and not detected at all times before $t_k$.
It is given by a formula similar to (\ref{Pkprod})
\begin{equation}\label{Pkprod2}
 P_k = p(t_k|t_{k-1})\prod_{j=1}^{k-1} \bar{p}(t_j|t_{j-1}) = p(t_k|t_{k-1}) \bar{P}_{k-1} .
\end{equation}
In the limit $\delta t\to 0$ we use (\ref{barPk}) and (\ref{w(t)2}) to write it as 
\begin{equation}\label{Pdens}
 {\cal P}(t) = w(t) e^{-\int_0^t dt'w(t')} ,
\end{equation}
where 
\begin{equation}\label{limit}
 {\cal P}(t)=\lim_{\delta t\to 0} \frac{P_k}{\delta t}
\end{equation}
is the probability density.
The result (\ref{Pdens}) coincides with that in \cite{jur-nik,nik_tajron}. 

Now we can determine the conditional probability of arrival at time $t_k=k\delta t$, if one knows that the particle 
has not yet arrived at the time $t_{k-1}=(k-1)\delta t$. In general,
this is the conditional probability $p(t_k|t_{k-1})$ given by (\ref{Pkprod2})
\begin{equation}\label{cond_final}
 p(t_k|t_{k-1})=\frac{P_k}{\bar{P}_{k-1}} .
\end{equation}
But the knowledge that the particle has not arrived is obtained by passive measurement, 
so $\bar{P}_{k-1}\equiv\bar{P}(t-\delta t)$ is given by the 3-dimensional version of 
(\ref{probkt2})
\begin{equation}\label{3DbarP}
 \bar{P}(t)=\int_{\bar{D}} d^3x\, |\psi({\bf x},t)|^2 . 
\end{equation}
Hence (\ref{cond_final}) is
\begin{equation}\label{cond_final2}
 p(t_k|t_{k-1})=\frac{P_k}{\displaystyle\int_{\bar{D}} d^3x\, |\psi({\bf x},t-\delta t)|^2} .
\end{equation}
The crucial thing here is the time-dependent denominator, which describes how the conditional probability 
of arrival changes with time $t$. If $\psi({\bf x},t)$ is a free wave packet that travels from $\bar{D}$
%{\em towards} the detector at $x=0$, 
towards the detector with the boundary $\partial\bar{D}$ (see Fig.~\ref{fig1}),
then the denominator decreases with $t$ (because less and less portion
of $\psi$ remains in $\bar{D}$), implying that the 
conditional probability of detection becomes {\em larger} with $t$.
This corresponds to avoidance of the quantum gambler's fallacy discussed in Sec.~\ref{SECqgambler}.

In principle, the probability density ${\cal P}(t)$ can be found from (\ref{Pdens}), but in practice 
it is rather hard. In \cite{jur-nik}, ${\cal P}(t)$ has been found in this way only for a very special case 
of a rectangular wave packet.
For a more general case, only an approximative expression for ${\cal P}(t)$ has been found
in \cite{nik_tajron}, with a rather complicated analysis.
In the following we find the exact and very general expression for ${\cal P}(t)$, 
with a much simpler analysis than in \cite{nik_tajron},
by a method that circumvents an explicit use of (\ref{Pdens}).

The main idea is to express everything in terms of $\bar{p}$ probabilities, rather than $p$ probabilities,
because $\bar{p}$ are probabilities that the particle is {\em not} detected, which 
correspond to passive measurements so are much easier to compute. Hence we write (\ref{Pkprod2}) as
\begin{eqnarray}\label{posP}
 P(t_k) &=& [1-\bar{p}(t_k|t_{k-1})] \prod_{j=1}^{k-1}\bar{p}(t_j|t_{j-1})
\nonumber \\
&=& \bar{P}(t_{k-1})-\bar{P}(t_{k}) ,
\end{eqnarray}
which can also be written as
\begin{equation}\label{diff}
 P(t)=-[\bar{P}(t)-\bar{P}(t-\delta t)] .
\end{equation}
Now we divide (\ref{diff}) with $\delta t$, take the limit $\delta t\to 0$ and use (\ref{limit}).  
This gives 
\begin{equation}\label{diff2}
 {\cal P}(t)=-\frac{d\bar{P}(t)}{dt} ,
\end{equation}
which coincides with the expression for ${\cal P}_{\rm arr}(t)$ in (\ref{arr_dep0}), as long as 
it is positive. Using (\ref{3DbarP}) and noting that it is the same as (\ref{barP(t)}),
we again obtain the result (\ref{Parr_fin})
\begin{equation}\label{Parr_fin_again}
 {\cal P}_{\rm arr}(t)=\left\{ 
\begin{array}{cl}
 \displaystyle \Phi_{\partial \bar{D}}(t) & \;\; {\rm for} \;\; \Phi_{\partial \bar{D}}(t)>0  \\
 0 & \;\; {\rm for} \;\; \Phi_{\partial \bar{D}}(t)\leq 0 . 
\end{array}
\right.
\end{equation}

We stress that (\ref{diff2}) is equivalent to (\ref{Pdens}), because both are derived from
(\ref{Pkprod2}). Indeed, if we write (\ref{barPk}) as
\begin{equation}\label{equiv} 
\bar{P}(t)=e^{-\int_0^t dt'w(t')}
\end{equation}
for $\delta t \to 0$, we see that
(\ref{Pdens}) can be derived directly from (\ref{diff2}).
However, for practical purposes, (\ref{diff2}) is more convenient than (\ref{Pdens}).
When one starts from (\ref{diff2}), then one is naturally inclined to use (\ref{3DbarP}),
which makes all calculations simple. One the other hand, when one starts from (\ref{Pdens}),
then one is naturally inclined to compute $w(t)$ from (\ref{w(t)2}), which makes the calculations 
complicated \cite{nik_tajron}. Nevertheless, the equivalence of the two approaches 
allows us to find a simple expression for $w(t)$.
First, we can use (\ref{equiv}) to write
\begin{equation}
w(t)=-\frac{d\bar{P}(t)/dt}{\bar{P}(t)}.
\end{equation}
Second, we can use (\ref{barP(t)}) and (\ref{FINAL}) to write it as
\begin{equation}\label{wfinal}
w(t)=\frac{\Phi_{\partial \bar{D}}(t)}{\displaystyle\int_{\bar{D}} d^3x\, |\psi({\bf x},t)|^2}.
\end{equation}
The denominator in (\ref{wfinal}) has the same probabilistic interpretation as that in
(\ref{cond_final2}).

\subsection{The origin of negative ``probabilities''}

Within the approach in Sec.~\ref{SECarrpass}, it is also illuminating to better 
understand how (\ref{diff2}), which is proportional to (\ref{posP}), can be negative 
when the flux is negative.
At first sight, it looks as if the first line in (\ref{posP}) cannot be negative, because it is a 
product of probabilities which cannot be negative. Probability, of course, cannot be negative, 
but a formula for computing probability can give negative numbers, in which case the formula is not applicable.
More specifically, the first factor $[1-\bar{p}(t_k|t_{k-1})]$ in (\ref{posP}) can become negative if 
$\bar{p}(t_k|t_{k-1})$ can become larger than 1, which can happen by the mechanism discussed 
in Sec.~\ref{SECzenobasic}. 

Let us show this in detail. We have 
\begin{equation}\label{negp1}
 \bar{p}(t_k|t_{k-1})=\bar{p}(t|t-\delta t)
=||\bar{\pi}e^{-iH\delta t}|\bar{\psi}_c(t-\delta t)\rangle||^2 ,
\end{equation}
where the conditional state $|\bar{\psi}_c(t-\delta t)\rangle \equiv |\bar{\psi}\rangle$ obeys 
\begin{equation}
 \bar{\pi}|\bar{\psi}\rangle=|\bar{\psi}\rangle .
\end{equation}
Thus for small $\delta t$, (\ref{negp1}) can be expanded as 
\begin{equation}\label{negp2}
 \bar{p}(t|t-\delta t)=1-i\delta t 
\left[ \langle\bar{\psi}|H|\bar{\psi}\rangle - \langle\bar{\psi}|H|\bar{\psi}\rangle^* \right] 
+ {\cal O}(\delta t^2) .
\end{equation}
Naively one would expect that the square bracket in (\ref{negp2}) vanishes, but actually it does not.
We have 
\begin{equation}
 \langle\bar{\psi}|H|\bar{\psi}\rangle = \int_{\bar{D}} d^3x\, \bar{\psi}^*H \bar{\psi}
= \alpha \int_{\bar{D}} d^3x\, \psi^*H \psi, 
\end{equation}
where $\alpha\equiv ||\chi_{\bar{D}}\psi||^{-2}$ and $\psi$ 
is the normalized solution of the Schr\"odinger equation $H\psi=i\partial_t\psi$.
The wave function $\psi({\bf x})$  
has a support everywhere on $\bar{D}\cup D$,
such that $\sqrt{\alpha}\psi({\bf x})=\bar{\psi}({\bf x})$ for ${\bf x}\in\bar{D}$.
In other words, $\alpha$ is a positive number originating from the fact that $\psi({\bf x})$
is normalized in  $\bar{D}\cup D$. 
Taking for the Hamiltonian
\begin{equation}\label{Hami}
 H=-\frac{\mbox{\boldmath $\nabla$}^2}{2m}+V({\bf x}) ,
\end{equation}
where $V({\bf x})$ is a real potential (describing the interactions in the detector region $D$, 
but possibly also some interactions in $\bar{D}$), we have
\begin{eqnarray}
 \langle\bar{\psi}|H|\bar{\psi}\rangle &=&
\alpha\int_{\bar{D}}d^3x\,\psi^*\left[-\frac{\mbox{\boldmath $\nabla$}^2}{2m}+V({\bf x}) \right] \psi  
\nonumber \\
&=& \frac{-\alpha}{2m}\int_{\bar{D}}d^3x\,\psi^*\mbox{\boldmath $\nabla$}^2 \psi + R_1
\nonumber \\
&=& \frac{-\alpha}{2m}\int_{\bar{D}}d^3x \left[ \mbox{\boldmath $\nabla$}(\psi^*\mbox{\boldmath $\nabla$} \psi)
 - ( \mbox{\boldmath $\nabla$} \psi^*)(\mbox{\boldmath $\nabla$} \psi) \right]    + R_1 
%\nonumber \\
%& & +\textcolor[rgb]{1,0,0}{{\rm real}} \textcolor[rgb]{0,0,1}{R}
\nonumber \\
&=& \frac{-\alpha}{2m}\int_{\partial\bar{D}}  d{\bf S}\cdot (\psi^*\mbox{\boldmath $\nabla$} \psi)
+ R_1 + R_2 ,
\end{eqnarray}
where $R_1\propto \int_{\bar{D}}d^3x\, \psi^*V\psi$ and 
$R_2\propto \int_{\bar{D}}d^3x\, (\mbox{\boldmath $\nabla$}\psi^*) (\mbox{\boldmath $\nabla$}\psi)$
%\begin{eqnarray}
%& R_1\propto \displaystyle\int_{\bar{D}}d^3x\, \psi^*V\psi \in\mathbb{R} , &
%\nonumber \\ 
%& R_2\propto \displaystyle\int_{\bar{D}}d^3x\, (\mbox{\boldmath $\nabla$}\psi^*) (\mbox{\boldmath $\nabla$}\psi) \in\mathbb{R} &
%\end{eqnarray}
are real, 
so they can be omitted because they do not contribute to (\ref{negp2}).
Likewise 
\begin{equation}
\langle\bar{\psi}|H|\bar{\psi}\rangle^*=
\frac{-\alpha}{2m}\int_{\partial\bar{D}}  d{\bf S}\cdot (\psi\mbox{\boldmath $\nabla$} \psi^*)
+ R_1 + R_2,
\end{equation}
so 
\begin{equation}
 \langle\bar{\psi}|H|\bar{\psi}\rangle - \langle\bar{\psi}|H|\bar{\psi}\rangle^*
= \frac{-\alpha}{2m}\int_{\partial\bar{D}}  d{\bf S}\cdot 
(\psi^* \!\stackrel{\leftrightarrow}{ \mbox{\boldmath $\nabla$} }\!\psi) .
\end{equation}
The usual quantum probability current associated with the Hamiltonian (\ref{Hami}) is 
\begin{equation}
{\bf j}=\frac{-i}{2m}\,\psi^* \!\stackrel{\leftrightarrow}{ \mbox{\boldmath $\nabla$} }\!\psi ,
\end{equation}
so we see that (\ref{negp2}) can finally be written as
\begin{equation}
\bar{p}(t|t-\delta t)=1-\delta t\, \alpha\int_{\partial\bar{D}}  d{\bf S}\cdot{\bf j}
=1-\delta t\,\alpha\,\Phi_{\partial\bar{D}}. 
\end{equation}
Thus we see that the formula for probability gives a value larger than 1 
precisely when the flux $\Phi_{\partial\bar{D}}$ is negative.

\section{Conclusion and outlook}
\label{SECdisc}

In this paper we have explored some surprising mathematical features of infinite dimensional Hilbert spaces
to develop a theoretical basis for a new kind of projective quantum measurement, 
called passive quantum measurement. Physically, passive quantum measurement is similar to a classical measurement,
in the sense that it can be interpreted us an update of information about the physical system without 
a physical change of the system itself. It is expected to be associated with negative measurement outcomes, 
for example when a particle detector does not click.   
An example is a quantum arrival experiment where, from the fact that the detector has not clicked yet,
we learn that the particle has not arrived yet. However, any such update of information corresponds to a projection
of the quantum state, and typical projections in quantum mechanics cannot be interpreted in such a passive way.
So, when the corresponding projector $\bar{\pi}$ does not commute with the Hamiltonian $H$ 
that governs the evolution of the system in the absence of measurement, 
it is very nontrivial that such a passive measurement can be compatible with quantum mechanics. 
We have found that the mathematical condition for such a compatibility is expressible as  
$\bar{\pi}[H,\bar{\pi}]=0$, which is only possible in the infinite dimensional Hilbert space.
Moreover, we have found that such a property is closely related to the surprising
possibility of having a non-zero imaginary part 
of the expectation value of the hermitian Hamiltonian, which is also 
only possible in the infinite dimensional Hilbert space. 

As a physical application, we have used all those results to find the arrival time probability density ${\cal P}_{\rm arr}(t)$,
for particles that arrive at the detector.
We have found that it is equal to the flux $\Phi_{\partial\bar{D}}(t)$ of the quantum probability current 
through the outer boundary $\partial\bar{D}$ of the detector, provided that $\Phi_{\partial\bar{D}}(t)>0$.
When $\Phi_{\partial\bar{D}}(t)<0$, we have used physical arguments to conclude that ${\cal P}_{\rm arr}(t)=0$ 
because then the particle departs, rather than arrives.

In this paper we have studied passive quantum measurements and related features in the contexts of quantum arrival time,
quantum Zeno effect and quantum gambler's fallacy. No doubt, passive quantum measurements can be of interest
in many other contexts that we have not explored yet, and not even thought of. Moreover, as a byproduct of our mathematical analysis,
we have found a mathematical possibility of a negative real part of the expectation value of a positive Hamiltonian,
the physical interpretation of which is yet to be found. Thus we believe that the results of this paper have a great potential
to open many new research directions in both theoretical and experimental physics.

\section*{Acknowledgments}

%This work was supported  
%by the Ministry of Science of the Republic of Croatia.
The authors are grateful to F. Giacosa, T. Maudlin, R. Tumulka and H. Wiseman for useful comments on the manuscript.
The work of T.J. is supported by Croatian Science
Foundation Project No. IP-2020-02-9614.

%\section*{Statements and Declarations}
% OBAVEZNO ZA EPJ+
%
%Data sharing not applicable to this article as no datasets were generated or analyzed during the current study 
%and article describes entirely theoretical research.

\end{document}